\documentclass[12pt]{article}
\begin{document}
\title{THE PLANCK SCALE UNDERPINNING FOR SPACE TIME}
\author{B.G. Sidharth\\
International Institute for Applicable Mathematics \& Information Sciences\\
Hyderabad (India) \& Udine (Italy)\\
B.M. Birla Science Centre, Adarsh Nagar, Hyderabad - 500 063 (India)}
\date{}
\maketitle
\begin{abstract}
We provide a rationale for the Planck scale being the minimum scale in the universe, as also its specific numerical values. In the process we answer the question of why the Planck scale is $10^{20}$ times the Compton scale of elementary particles. These considerations show how the Planck scale provides an underpinning for space time.
\end{abstract}
\section{Introduction}
In the latest approaches such as Quantum Superstrings (or M-Theory) or Quantum Gravity Theories it is generally accepted that the Planck scale defines a minimum scale for the universe \cite{ven,kempf,garay,crane,psu,psp}. In these schemes there is a maximal mass, the Planck mass $m_P \sim 10^{-5}gms$ which is defined by
\begin{equation}
m_P = \left(\frac{\hbar c}{G}\right)^{1/2} \approx 10^{-5}gms\label{e1}
\end{equation}
Using the value for $m_P$ we can define the Planck length $l_P \sim 10^{-33}cms$ and the Planck time $t_P \sim 10^{-42}secs$, which are the Compton lengths and times for the mass in (\ref{e1}). It may be mentioned that these values were postulated by Planck himself. Today the values for the minimum scale as given in (\ref{e1}) are taken for granted. We first provide a rationale for the numerical value of the Planck scale.\\
\section{The Planck Scale}
Our starting point is the model for the underpinning at the Planck scale for the universe. This is a collection of $N$ Planck scale oscillators (Cf.refs.\cite{psu,psp,uof,gip,ng} for details). We do not need to specfify $N$. We have in this case the following well known relations
$$R = \sqrt{N}l, Kl^2 = kT,$$
\begin{equation}
\omega^2_{max} = \frac{K}{m} = \frac{kT}{ml^2}\label{e2}
\end{equation}
In (\ref{e2}), $R$ is of the order of the diameter of the universe, $K$ is the analogue of the spring constant, $T$ is the effective temperature while $l$ is the analogue of the Planck length, $m$ the analogue of the Planck mass and $\omega_{max}$ is the frequency-the reason for the subscript $max$ will be seen below. We do not yet give $l$ and $m$ their usual values as given in (\ref{e1}) for example, but rather try to deduce these values.\\
We now use the well known result that the individual minimal oscillator particles are black holes or mini universes as shown by Rosen \cite{rosen}. So using the well known Beckenstein temperature formula for these primordial black holes \cite{ruffini}, that is
$$kT = \frac{\hbar c^3}{8\pi Gm}$$
in (\ref{e2}) we get,
\begin{equation}
Gm^2 \sim \hbar c\label{e3}
\end{equation}
We can easily verify that (\ref{e3}) leads to the value $m \sim 10^{-5}gms$. In deducing (\ref{e3}) we have used the typical expressions for the frequency as the inverse of the time - the Compton time in this case and similarly the expression for the Compton length. However it must be reiterated that no specific values for $l$ or $m$ were considered in the deduction of (\ref{e3}).\\
We now make two interesting comments. Cercignani and co-workers have shown \cite{cer1,cer2} that when the gravitational energy becomes of the order of the electromagnetic energy in the case of the Zero Point oscillators, that is
\begin{equation}
\frac{G\hbar^2 \omega^3}{c^5} \sim \hbar \omega\label{e4}
\end{equation}
then this defines a threshold frequency $\omega_{max}$ above in which the oscillations become chaotic.\\
Secondly from the parallel but unrelated theory of phonons \cite{huang,rief}, which are also bosonic oscillators, we deduce a maximal frequency given by
\begin{equation}
\omega^2_{max} = \frac{c^2}{l^2}\label{e5}
\end{equation}
In (\ref{e5}) $c$ is, in the particular case of phonons, the velocity of propagation, that is the velocity of sound, whereas in our case this velocity is that of light. Frequencies greater than $\omega_{max}$ in (\ref{e5}) are meaningless.  We can easily verify that (\ref{e4}) and (\ref{e5}) give back (\ref{e3}).\\
Finally we can see from (\ref{e2}) that, given the value of $l_P$ and using the value of the radius of the universe, viz., $R \sim 10^{27}$, we can deduce that, 
\begin{equation}
N' \sim 10^{120}\label{e6}
\end{equation}
\section{The Gauge Hierarchy Problem}
We next consider a long standing puzzle, the so called gauge hierarchy problem, that is why is there such a wide gap between the mass of a Planck particle, $10^{-5}gms$ and the mass of a typical elementary particle $\sim 10^{-25}gms$. We show that the answer to this problem lies in a particular characterization of gravitation. This moreover also provides a picture of a Planck scale underpinning for the entire universe itself.\\
In 1997 the author put forward a model in which particles are created fluctuationally in a phase transition from the background Zero Point Field or Dark Energy. This model lead to dramatic consequences: The Universe would be accelerating and expanding with a small cosmological constant. Besides, several longstanding puzzling relations of the so called large number genre which had no explanation whatsoever, were now deduceable from the theory.\\
At that time the accepted Standard Big Bang model predicted a dark matter filled decelerating universe--in other words the exact opposite. In 1998 however the first results were announced by Perlmutter, Kirshner and coworkers, from a study of distant supernovae that the universe was indeed accelerating and expanding with a small cosmological constant, contrary to belief. These conclusions were subsequently reconfirmed several times. This work was the breakthrough of the year 1998 of the American Association for the Advancement of Science \cite{science} while Dark Energy itself was subsequently confirmed through the WMAP and the Sloan Digital Sky Survey. This in fact was the breakthrough of the year 2003 \cite{science2}.\\
To recapitulate the author's model \cite{mg8,ijmpa,ijtp,csfnc,csfnc2,cu,uof}, we give a simple picture and will return to the nuances later.\\
Our starting point is the all permeating Zero Point Field or the Dark Energy, from which the elementary particles are created. As Wheeler put it \cite{mwt}, ``From the zero-point fluctuations of a single oscillator to the fluctuations of the electromagnetic field to geometrodynamic fluctuations is a natural 
order of progression...''\\ Let us consider,
following Wheeler a harmonic oscillator in its ground state. The
probability amplitude is
$$\psi (x) = \left(\frac{m\omega}{\pi \hbar}\right)^{1/4}
e^{-(m\omega/2\hbar)x^2}$$ for displacement by the distance $x$ from
its position of classical equilibrium. So the oscillator fluctuates
over an interval
$$\Delta x \sim (\hbar/m\omega)^{1/2}$$ The electromagnetic field is
an infinite collection of independent oscillators, with amplitudes
$X_1,X_2$ etc. The probability for the various oscillators to have
emplitudes $X_1, X_2$ and so on is the product of individual
oscillator amplitudes:
$$\psi (X_1,X_2,\cdots ) = exp [-(X^2_1 + X^2_2 + \cdots)]$$ wherein
there would be a suitable normalization factor. This expression gives
the probability amplitude $\psi$ for a configuration $B (x,y,z)$ of
the magnetic field that is described by the Fourier coefficients
$X_1,X_2,\cdots$ or directly in terms of the magnetic field
configuration itself by
$$\psi (B(x,y,z)) = P exp \left(-\int \int \frac{\bf{B}(x_1)\cdot
\bf{B}(x_2)}{16\pi^3\hbar cr^2_{12}} d^3x_1 d^3x_2\right).$$ 
$P$ being a normalization factor. Let us consider a configuration where the
magnetic field is everywhere zero except in a region of dimension $l$,
where it is of the order of $\sim \Delta B$. The probability amplitude
for this configuration would be proportional to
$$\exp [-(\Delta B)^2 l^4/\hbar c)$$ 
So the energy of fluctuation in a region of length $l$ is given by finally \cite{mwt,r24,r25}
$$B^2 \sim \frac{\hbar c}{l^4}$$
In the above if $l$ is taken to be
the Compton wavelength of a typical elementary  particle, then we
recover its energy $mc^2$, as can be easily verified.\\
It may be mentioned
that Einstein himself had believed that the electron was a result
 of
such a condensation from the background electromagnetic field
(Cf.\cite{r26,cu} for details). We will return to this point
again. We also take the pion to represent a typical elementary
particle, as in the literature.\\
 
To proceed, as there are $N \sim 10^{80}$ such particles in the universe, we get
\begin{equation}
Nm = M\label{e1a}
\end{equation}
where $M$ is the mass of the universe.\\
 
In the following we will use $N$ as the sole cosmological parameter.\\
 
Equating the
gravitational potential energy of the pion in a three dimensional
isotropic
 sphere of pions of radius $R$, the radius of the universe,
with the rest
 energy of the pion, we can deduce the well known
relation \cite{csfnc,r28,r29}
\begin{equation}
R \approx \frac{GM}{c^2}\label{e2a}
\end{equation}
where $M$ can be obtained from (\ref{e1a}).\\
 
We now use the fact
that given $N$ particles, the fluctuation in the particle number is of
the
 order $\sqrt{N}$\cite{r29,huang,ijmpa,ijtp,r5,mg8}, while a typical time
interval for the
 fluctuations is $\sim \hbar/mc^2$, the Compton
time. We will come back to
 this point later in the context
of the minimum Planck scale: Particles are created and destroyed - but
the ultimate result is that $\sqrt{N}$ particles are created. So we
have, as we saw briefly earlier,
\begin{equation}
\frac{dN}{dt} = \frac{\sqrt{N}}{\tau}\label{ex}
\end{equation}
whence on integration we get, (remembering that we are almost in the
continuum region),
\begin{equation}
T = \frac{\hbar}{mc^2} \sqrt{N}\label{e3a}
\end{equation}
We can easily verify that the equation is
indeed satisfied
 where $T$ is the age of the universe. Next by
differentiating (\ref{e2a}) with
 respect to $t$ we get
\begin{equation}
\frac{dR}{dt} \approx HR\label{e4a}
\end{equation}
where $H$ in (\ref{e4a}) can be identified with the Hubble Constant,
and using
 (\ref{e2a}) is given by,
\begin{equation}
H = \frac{Gm^3c}{\hbar^2}\label{e5a}
\end{equation}
Equation (\ref{e1a}), (\ref{e2a}) and (\ref{e3a}) show that in this
formulation, the correct mass, radius, Hubble constant and age of
the universe can be deduced given $N$ as the sole cosmological or
large scale parameter. Equation (\ref{e5a}) can be written as
\begin{equation}
m \approx \left(\frac{H\hbar^2}{Gc}\right)^{\frac{1}{3}}\label{e6a}
\end{equation}
Equation (\ref{e6a}) has been empirically known as an "accidental" or
"mysterious" relation. As observed by Weinberg\cite{r10}, this is
unexplained: it relates a single cosmological parameter $H$ to
constants from microphysics. We will touch upon this micro-macro
nexus again. In our formulation, equation (\ref{e6a}) is no longer a
mysterious coincidence but
 rather a consequence.\\
 
As (\ref{e5a})
and (\ref{e4a}) are not exact equations but rather,
 order of
magnitude relations, it follows, on differentiating (\ref{e4a}) that
a
 small cosmological constant $\wedge$ is allowed such that
$$\wedge < 0 (H^2)$$
 This is consistent with observation and shows
that $\wedge$ is very small -- this has been a puzzle,
 the so called
cosmological constant problem alluded to, because in conventional
theory, it turns out to be huge \cite{rj}. Some $10^{70}$ times higher in fact! But it poses no problem in
this formulation.\\
 
To proceed we observe that because of the fluctuation of
$\sim \sqrt{N}$ (due to the ZPF),
 there is an excess electrical
potential energy of the electron, which infact
 we have identified as
its inertial energy. That is \cite{ijmpa,r29},
$$\sqrt{N} e^2/R \approx mc^2.$$
 On using (\ref{e2a}) in the above,
we recover the well known Gravitation-electromagnetism
 ratio viz.,
\begin{equation}
e^2/Gm^2 \sim \sqrt{N} \approx 10^{40}\label{e7}
\end{equation}
or without using (\ref{e2a}), we get, instead, the well known so
called
 Weyl-Eddington formula,
\begin{equation}
R = \sqrt{N}l\label{e8}
\end{equation}
(It appears that this was first noticed by H. Weyl \cite{r31}).
Infact (\ref{e8}) is the spatial counterpart of (\ref{e3a}). If we
combine (\ref{e8}) and (\ref{e2}), we get,
\begin{equation}
\frac{Gm}{lc^2} = \frac{1}{\sqrt{N}} \propto T^{-1}\label{e9}
\end{equation}
where in (\ref{e9}), we have used (\ref{e3a}). Following Dirac (cf.also
\cite{r32})
 we treat $G$ as the variable, rather than the quantities
$m, l, c \,\mbox{and}\,
 \hbar$ (which we will call micro physical
constants) because of their central role
 in atomic (and sub atomic)
physics.\\
 Next if we use $G$ from (\ref{e9}) in (\ref{e5a}), we can
see that
\begin{equation}
H = \frac{c}{l} \quad \frac{1}{\sqrt{N}}\label{e10}
\end{equation}
Thus apart from the fact that $H$ has the same inverse time dependance
on
 $T$ as $G$, (\ref{e10}) shows that given the microphysical
constants, and
 $N$, we can deduce the Hubble Constant also, as from
(\ref{e10}) or (\ref{e5a}).\\
 Using (\ref{e1a}) and (\ref{e2a}), we can
now deduce that
\begin{equation}
\rho \approx \frac{m}{l^3} \quad \frac{1}{\sqrt{N}}\label{e11}
\end{equation}
Next (\ref{e8}) and (\ref{e3a}) give,
\begin{equation}
R = cT\label{e12}
\end{equation}
(\ref{e11}) and (\ref{e12}) are consistent with observation.\\
With regard to the time variation of $G$, the issue is debatable and model dependent. Measurements on the earth and of the planets, and perhaps most accurate of all, Pulsars indicates a value $\sim 10^{-10}$, though values $10^{-11}$ and $10^{-12}$ have also appeared in some studies \cite{mel,uzan}.\\
We return to the question of why the Planck mass is some $10^{20}$ times the mass of an elementary particle, for example pions or Protons or Electrons (in the large number sense).\\
It is well known that the Planck mass is defined by (\ref{e1}) \cite{mwt,cu}
Alternatively the Planck mass defines a black hole having the Schwarzchild radius given by
\begin{equation}
\frac{2Gm_P}{c^2} \sim l_P \, \sim 10^{-33}cm\label{e2aa}
\end{equation}
In (\ref{e2aa}) $l_P$ is the Planck length.\\
While $m_P$ is $\sim 10^{-5}gms$, a typical elementary particle has a mass $m \sim 10^{-25}gms$. (As mentioned in this order of magnitude sense it does not make much difference, if the elementary particle is an electron or pion or proton (Cf.ref.\cite{mwt})).\\
We now recall that as already shown we have \cite{ijmpa,ijtp,fpl,uof}
\begin{equation}
G = \frac{\hbar c}{m^2\sqrt{N}}\label{e3aa}
\end{equation}
In (\ref{e3aa}) $N \sim 10^{80}$ is the well known number of elementary particles in the universe, which features in the Weyl-Eddington relations as also the Dirac Cosmology.\\
What is interesting about (\ref{e3aa}) is that it shows gravitation as a distributional effect over all the $N$ particles in the universe \cite{fpl,uof}.\\
Let us rewrite (\ref{e2aa}) in the form
\begin{equation}
G \approx \frac{\hbar c}{m^2_P}\label{e4aa}
\end{equation}
remembering that the Planck length is also the Compton length of the Planck mass. (Interestingly an equation like (\ref{e2aa}) or (\ref{e4aa}) also follows from Sakharov's treatment of gravitation \cite{sakharov}.) A division of (\ref{e3}) and (\ref{e4aa}) yields
\begin{equation}
m^2_P = \sqrt{N} m^2\label{e5aa}
\end{equation}
Equation (\ref{e5aa}) immediately gives the ratio $\sim 10^{20}$ between the Planck mass and the mass of an elementary particle.\\
It is interesting that in (\ref{e3aa}) if we take $N \sim 1$, then we recover (\ref{e4aa}). So while the Planck mass in the spirit of Rosen's isolated universe and the Schwarzchild black hole uses the gravitational interaction in isolation, as seen from (\ref{e3aa}), elementary particles are involved in the gravitational interaction with all the remaining particles in the universe.\\
Finally rememebring that $Gm_P^2 \sim e^2$, as can also be seen from (\ref{e4aa}), we get from (\ref{e3aa})
\begin{equation}
\frac{e^2}{Gm^2} \sim \frac{1}{\sqrt{\bar{N}}}\label{e6aa}
\end{equation}
Equation (\ref{e6aa}) is the otherwise empirically well known electromagnetism-gravitation coupling constant ratio, but here it is deduced from the theory.\\
It may be remarked that one could attempt an explanation of (\ref{e5aa}) from the point of view of SuperSymmetry or Brane theory, but these latter have as yet no experimental validation \cite{gor}.
\section{The Universe as Planck Oscillators}
What we have tried to argue is that a typical elementary particle like a pion could be considered to be the result of
$n \sim 10^{40}$ evanescent Planck scale particles. The argument was based on 
random motions and also on the modification to the Uncertainity Principle.
We will now consider the problem from a different point of view,
which not only reconfirms the above result, but also enables an elegant
extension to the case of the entire universe itself.
Let us consider an array of $N$ particles, spaced a distance $\Delta x$
apart, which behave like oscillators, that is as if they were connected by
springs. We then have as seen
\begin{equation}
r  = \sqrt{N \Delta x^2}\label{e1d}
\end{equation}
\begin{equation}
ka^2 \equiv k \Delta x^2 = \frac{1}{2}  k_B T\label{e2d}
\end{equation}
where $k_B$ is the Boltzmann constant, $T$ the temperature, $r$ the extent  and $k$ is the 
spring constant given by
\begin{equation}
\omega_0^2 = \frac{k}{m}\label{e3d}
\end{equation}
\begin{equation}
\omega = \left(\frac{k}{m}a^2\right)^{\frac{1}{2}} \frac{1}{r} = \omega_0
\frac{a}{r}\label{e4d}
\end{equation}
We now identify the particles with Planck masses, set $\Delta x \equiv a = 
l_P$, the Planck length. It may be immediately observed that use of 
(\ref{e3d}) and (\ref{e2d}) gives $k_B T \sim m_P c^2$, which ofcourse agrees 
with the temperature of a black hole of Planck mass. Indeed, as noted, Rosen had shown that a Planck mass particle at the Planck scale  can be considered to be a
universe in itself. We also use the fact alluded to that  a typical elementary particle
like the pion can be considered to be the result of $n \sim 10^{40}$ Planck
masses. Using this in (\ref{e1d}), we get $r \sim l$, the pion
Compton wavelength as required. Further, in this latter case, using (48) and the fact that $N = n \sim 10^{40}$, and (\ref{e2d}),i.e. $k_BT = kl^2/N$ and  (\ref{e3d}) and
(\ref{e4d}), we get for a pion, remembering that $m^2_P/n = m^2,$ 
$$k_ B T = \frac{m^3 c^4 l^2}{\hbar^2} = mc^2,$$
which of course is the well known formula for the Hagedorn temperature for
elementary particles like pions. In other words, this confirms the conclusions
in the previous section, that we can treat an elementary particle as a series of some
$10^{40}$ Planck mass oscillators. However it must be observed from 
(\ref{e2d}) and (\ref{e3d}), that while the Planck mass gives the highest
energy state, an elementary particle like the pion is in the lowest energy
state. This explains why we encounter elementary particles, rather than
Planck mass particles in nature. Infact as already noted \cite{cu}, a Planck
mass particle decays via the Bekenstein radiation within a Planck time
$\sim 10^{-42}secs$. On the other hand, the lifetime of an elementary particle
would be very much higher.\\
In any case the efficacy of our above oscillator model can be seen by the fact that we recover correctly the masses and Compton scales in the order of magnitude sense and also get the correct Bekenstein and Hagedorn formulas as seen above, and get the correct estimate of the mass of the universe itself, as will be seen below.\\
Using the fact that the universe consists of $N \sim 10^{80}$ elementary
particles like the pions, the question is, can we think of the universe as
a collection of $n N \, \mbox{or}\, 10^{120}$ Planck mass oscillators? We directly deduced this value a little earlier, in fact. This is what we will now
show. Infact if we use equation (\ref{e1d}) with
$$\bar N \sim 10^{120},$$
we can see that the extent $r \sim 10^{28}cms$ which is of the order of the diameter of the
universe itself. Next using (\ref{e4d}) we get
\begin{equation}
\hbar \omega_0^{(min)} \langle \frac{l_P}{10^{28}} \rangle^{-1} \approx m_P c^2 \times 10^{60} \approx Mc^2\label{e5d}
\end{equation}
which gives the correct mass $M$, of the universe which in contrast to the earlier pion case, is the highest energy state while the Planck oscillators individually are this time the lowest in this description. In other words the universe
itself can be considered to be described in terms of normal modes of Planck scale oscillators.\\ 
The above gives a rationale for the figure $10^{120}$ Planck oscillators which is derived from the observed $10^{80}$ elementary particles in the universe and considerations in Section 2.\\
3. In fact it has already been shown that the universe has been considered to be a coherent state of $N'$ Planck oscillators, where $M$ is given by (\ref{e6}). In fact if we use (\ref{e2}) we can deduce that
\begin{equation}
M = \sqrt{N'} m_P\label{ey}
\end{equation}
Further, using considerations in point 2 we can also deduce that
$$m = m_P /\sqrt{n}$$
\begin{equation}
l = l_P \sqrt{n}\label{ez}
\end{equation}
$$\tau = \tau_P \sqrt{n}$$
where $m, l, \tau$ are the mass, the Compton wavelength and Compton time of a typical elementary particle like the pion. This brings us to the question: While the Planck scale may provide the underpinning, in real life it is the Compton scale of elementary particles that we encounter. How do we make this transition. We consider below some other approaches.
\section{Other Approaches}
1. It is well known that the energy of the Zero Point Field is given by
\begin{equation}
(\Delta B)^2 \geq \hbar c/L^4\label{g1}
\end{equation}
If in (\ref{g1}) $L$ the extent over which the energy is spread, is taken to be the Compton wavelength, we recover the mass of the elementary particle. Alternatively it is known that the spectral density of the vacuum field is given by
\begin{equation}
\rho (\omega) = const. \omega^3\label{g2}
\end{equation}
where Lorentz invariance requires that the constant is given by
$$\frac{\hbar}{2\pi^2 c^3}$$
In conventional theory this leads to an unacceptably high energy density of $10^{15}gms$ per $cm^3$, but at the Compton wavelength, we get back the above result.\\
2. In a similar vein, we can argue that when the Zero Point Field is taken into account, at the Compton wavelength scale the momentum coordinate commutators which otherwise vanish in Classical Theory go over to the Quantum Mechanical commutators \cite{bgsdark}. From this point of view the Zero Point Field, as argued by some protogonists of Stochastic Electrodynamics, gives rise to Quantum behavior.\\
3. What we have seen above is that from the background Zero Point Field, Plank scale oscillators condense out. Let us suppose that $n$ such particles are formed. We can then use the well known fact that for a collection of ultra relativistic particles, in this case the Planck oscillators, the various centres of mass form a two dimensional disk of radius $l$ given by
\begin{equation}
l \approx \frac{\beta}{m_e c}\label{ex17}
\end{equation}
where in (\ref{ex17}) $m_e (\approx m$ in the large number sense) is the electron mass and  $\beta$ is the angular momentum of the system. Further $l$ is such that for distances $r < l$, we encounter negative energies (exactly as for the Compton length). It will at once be apparent that for an electron, for which $\beta = \frac{\hbar}{2}$, (\ref{ex17}) gives the Compton wavelength. We can further characterize (\ref{ex17}) as follows: By the definition of the angular momentum of the system of Planck particles moving with relativistic speeds, we have
\begin{equation}
\frac{\hbar}{2} = m_P c \int^l_0 r^2 drd \Theta \, \sim m_P c \sigma l^3 = m_e cl\label{ex18}
\end{equation}
In (\ref{ex18}) we have used the fact that the disk of mass centres is two dimensional, and $\sigma$ has been inserted to stress the fact that we are dealing with a two dimensional density, so that $\sigma$ while being unity has the dimension
$$\left[\frac{1}{L^2}\right]$$
The right side of (\ref{ex18}) gives the angular momentum for the electron.
From (\ref{ex18}) we get
\begin{equation}
\sigma l^2 m_P = m_e\label{ex19}
\end{equation}
which ofcourse is correct.\\
Alternatively from (\ref{ex19}) we can recover $n \sim 10^{40}$, in the large number sense.\\ 
All this shows how the Compton scale of elementary particles emerges from the Planck scale.
\section{Concluding Remarks}
The universe which we encounter is at the Compton scale of elementary particles. The various elementary particles which constitute the universe are in a sense incoherent in that they can be treated as independent particles which are not coupled or linked. Yet they occupy a single space time, remembering that at large distances the various particles interact via the relatively weak force of gravitation. This is expressed by the relation
$$M = Nm,$$
where $N$ is the number of particles in the universe and $m$ a typical elementary particle mass, $M$ being the mass of the universe.\\
However at a higher energy or smaller scale of observation, viz., the Planck scale, the universe is seen as coherent collection of $N' \sim 10^{120}$ Planck oscillators, which in fact provides the underpinning for all of space time. In this case we have,
$$M = \sqrt{N'} m_P$$
where we are now speaking of the Planck mass. As described elsewhere (Cf.refs.\cite{uof,gip}), we can describe the above difference in the following manner: The most fundamental scale is the Planck scale, represened by wave functions $\psi_\imath$. However by the Random Phase axiom, the superposition of the $\psi_\imath$s reduces to the simpler superposition of $\phi_\imath$s, these latter representing the wave functions at the Compton scale of elementary particles.

\end{document}